\documentclass{icrc}

\usepackage{times}
\usepackage{graphicx} 

\begin{document}

\title{Pachmarhi Array of \v Cerenkov Telescopes and its Sensitivity}
\author[1]{V. R. Chitnis}
\author[1]{B. S. Acharya}
\author[1]{P. N. Bhat}
\author[1]{K. S. Gothe}
\author[1]{A. V. John}
\author[1]{P. Majumdar}
\author[1]{B. K. Nagesh}
\author[1]{M. A. Rahman}
\author[1]{B. B. Singh}
\author[1]{S. S. Upadhyaya}
\author[1]{B. L. V. Murthy}
\author[1]{P. R. Vishwanath}
\affil[1]{Tata Institute of Fundamental Research, Homi Bhabha Road,
Mumbai 400 005, India}

\correspondence{V. R. Chitnis (vchitnis@tifr.res.in)}

\runninghead{Chitnis {\it et al.}: Sensitivity of PACT}
\firstpage{1}
\pubyear{2001}


\maketitle

\begin{abstract}
Pachmarhi Array of \v Cerenkov Telescopes (PACT) has been designed to 
search for celestial TeV $\gamma-$rays using the wavefront sampling
technique. PACT, located at Pachmarhi, (latitude 22$^\circ$ 28$^\prime$
N, longitude 76$^\circ$ 26$^\prime$ E, altitude 1075 m) consists of
25 telescopes  deployed over an area of 80 m $\times$ 100m. Each
telescopes consists of 7 parabolic reflectors, each viewed by a fast 
phototube behind a 3$^\circ$ mask at the focus. The density and the 
arrival time of the photons at the PMT are recorded for each shower.
The energy threshold and collection area of the array are estimated,
from Monte Carlo simulations, to be $\sim$ 900 GeV and 10$^5$ m$^2$ 
respectively. The accuracy in determination of arrival angle of a
shower is estimated to be about 0.1$^\circ$ in the near vertical direction.
About 99\% of the off-axis hadronic events could be rejected from
directional information alone. Further, at least 75\% of the on-axis
hadronic events could be rejected using species sensitive parameters
derived from timing and density measurements. These cuts on data to reject
background would retain $\sim$ 44\% of the $\gamma-$ray signal. The
sensitivity of the array for a 5$\sigma$ detection of $\gamma-$ray
signal at a threshold energy of 1 TeV has been estimated to be 
$\sim$ 4.1 $\times$ 10$^{-12}$ photons cm$^{-2}$ s$^{-1}$ for an on
source exposure of 50 hours. The PACT set-up has been fully commissioned
and is collecting data. The details of the system parameters and sensitivity
will be presented.
\end{abstract}

\section{Introduction}

Ground based atmospheric \v Cerenkov technique is, at present, the only way 
by which VHE $\gamma-$rays from astronomical objects are detected. Using this 
technique TeV $\gamma-$rays have been detected successfully from a number of 
galactic sources including pulsars, supernova remnants etc as well as from 
extra-galactic objects which are AGNs of blazar class. There are a number 
experiments based on this technique operating across the globe using either 
angular imaging technique or wavefront sampling technique. These are two 
complementary ways of studying the \v Cerenkov emission from air showers 
generated by $\gamma-$rays from astronomical sources. Imaging technique has 
been successfully exploited by several experiments including Whipple, CAT, 
CANGAROO, HEGRA, TACTIC etc. On the other hand, the experiments like 
CELESTE, STACEE, SOLAR-2, GRAAL and PACT are based on wavefront sampling 
technique. These experiments consist of an array of \v Cerenkov telescopes 
which sample the \v Cerenkov pool at the observation level. These experiments 
measure the arrival time of \v Cerenkov shower front and \v Cerenkov photon 
density at various locations in the \v Cerenkov pool. In this paper we 
discuss some of the design aspects and performance parameters of Pachmarhi 
Array of \v Cerenkov Telescopes or PACT, which is based on wavefront sampling 
technique.

\section{PACT : Instrument Details}

Pachmarhi Array of \v Cerenkov Telescopes is located at Pachmarhi (latitude 
22$^\circ$ 28$^\prime$ N, longitude 76$^\circ$ 26$^\prime$ E, altitude 1075 
m) in Central India. This array consists of 5 $\times$ 5 array of 25 \v 
Cerenkov telescopes spread over a rectangular area of 80 $m$ $\times$ 100 $m$ 
(see Figure 1). Spacing between the telescopes is 20 $m$ in E-W direction 
and 25 $m$ in N-S direction. Each telescope consists of seven para-axially 
mounted parabolic reflectors of diameter 0.9 $m$ each with $f/d$ ratio being 
$\sim$ 1 (see Figure 2). These reflectors are fabricated indigenously and 
their optical quality is such that the size of the image of a point source 
is $\le$ 1$^\circ$. They are back-coated and their reflectivity in visible 
range is $\sim$ 70\%. These reflectors are mounted in hexagonal pattern and 
total reflector area per telescope is $\sim$ 4.45 m$^2$. A fast phototube of 
the type EMI9807B is mounted at the focus of each reflector. Field of view 
defined by the photo-cathode diameter is $\sim$ 3$^\circ$ FWHM. 



Telescopes are equatorially mounted and each telescope is independently
steerable in both E-W and N-S direction within $\pm$45$^\circ$.  The 
movement of telescopes is remotely controlled by Automatic Computerized 
Telescope Orientation System (ACTOS). The hardware consists of a 
semi-intelligent closed loop stepper motor system which senses the angular 
position using a gravity based transducer called clinometer with an accuracy 
of 1$^\prime$. The two clinometers, one each in N-S and E-W direction, are 
accurately calibrated using stars. The system can orient to the putative 
source with an accuracy of $\sim$ (0.003 $\pm$ 0.2). The source pointing 
is monitored at an accuracy of $\sim$0.05$^\circ$ and corrected in real time 
whenever the error exceeds 0.05$^\circ$.

High voltages to individual phototubes are applied through a computerized
control system (called CARAMS, Computerized Automated Rate Adjustment and 
Monitoring System) and voltages as well as count rates from individual 
phototubes are continuously monitored. Array is divided into four sub-groups
or sectors of six telescopes each for data acquisition. At the centre of
each sector there is a field signal processing centre (FSPC). Pulses from 
phototubes are brought to the respective stations using low attenuation 
coaxial cables of the type RG213, of length $\sim$ 40 $m$. These pulses are 
processed by front end electronics in the FSPC and informations such as \v 
Cerenkov photon density (ADC) and arrival time of \v Cerenkov shower front 
at the mirror given by TDC (resolution 0.25 $ns$) and event arrival time 
(UTC) correct to 1 $\mu s$ are recorded. The ADC and TDC data are recorded 
for six peripheral mirrors of each telescope. In addition to this, some 
information relevant to the entire array is recorded by master signal 
processing centre (MSPC) in the control room at the centre of the array. This 
information includes relative arrival time of \v Cerenkov shower front at 
individual telescopes, absolute arrival time of the shower front accurate 
to $\mu s$ derived from a real time clock. The real time clocks (RTC) in 
FSPC's and MSPC's are synchronized with each other and with a GPS clock. 
Using RTC information events recorded in individual FSPC's are collated 
with those recorded in MSPC off-line. Data recording in MSPC as well as in 
FSPC is carried out using networked Linux based system. Details of the data 
acquisition system are given elsewhere (Bhat {\it et al.}, 2001a). This 
array is fully operational since December, 2000. Preliminary results 
obtained from observations carried out by this array on sources including 
crab nebula and mkn 421 are discussed elsewhere (Vishwanath {\it et al.}, 
2001 and Bhat {\it et al.}, 2001b).

\section{Energy threshold of PACT}

The night sky background (NSB) is 
the limiting factor in detecting \v Cerenkov photons from low energy 
primaries and this decides the low energy threshold of the experiment. The 
NSB measured at Pachmarhi, over the range of the spectral response of the 
phototube, is $\sim$ 3.3 $\times$ 10$^8$ $ph~cm^{-2}~s^{-1}$.

\begin{figure}[t]
\includegraphics[width=8.3cm]{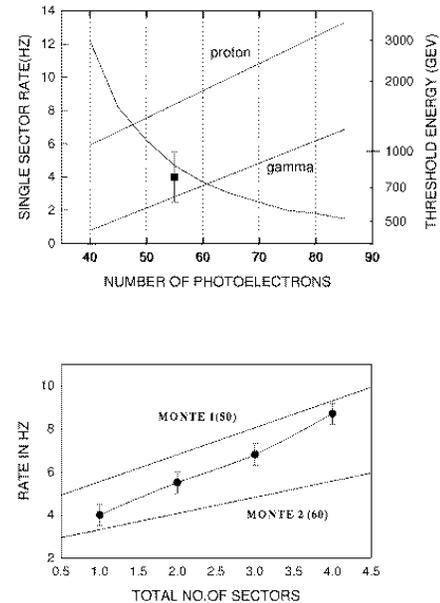}
\caption{Upper panel shows single sector trigger rate $vs$ no. of
photo-electrons per telescope. Curve corresponds to simulated data. 
Observed trigger rate is indicated by a point with error bar. Also shown 
in figure is the threshold energy (scale on right) as a function of number 
of photo-electrons per telescope for $\gamma-$rays and protons. Lower panel 
shows the observed trigger rate $vs$ number of sectors. Also shown are the 
expected trigger rates with photo-electron threshold of 50 and 60 per 
telescope, based on simulations.}
\end{figure}

In order to estimate the expected performance of the array, large number 
of $\gamma-$ray and proton showers are simulated taking into account 
various design features of the array. For $\gamma-$ray primaries, energies 
are chosen from a power law spectrum with a slope of -1.4 over the range of 
500 GeV to 20 TeV. Whereas for protons, slope of the spectrum is -1.66. 
Factors like atmospheric attenuation, reflectivity of mirrors, quantum 
efficiency of phototubes, attenuation in cables etc. are taken into 
account in simulations. Same trigger criteria as used in the experiment are
applied to the simulated events. In the experiment, the seven individual 
mirror outputs are added to get the analog sum called the `royal sum' for 
each telescope. Each sector has six royal sum pulses corresponding to six 
telescopes. The royal sums are discriminated to yield a counting rate of 
$\sim 40~kHz$. A trigger is generated when at least four out of six royal 
sums are present. From the simulated data, the trigger rates were obtained 
for each sector for various photo-electron  thresholds ranging from 35 to 
100. The variation of trigger rate as a function of photo-electron threshold 
is shown in Figure 3 (upper panel). The observed trigger rate is also shown. 
It can be seen that the trigger rate corresponds to the threshold of about 
55 photo-electrons per telescope. 

Lower panel of Figure 3 shows the experimental trigger rates when the 
number of sectors increased from 1 to 4. The overall trigger rate essentially
varies as the square root of total mirror area. It  increases from about 4 Hz 
for a one sector to 9 Hz when all the four sectors are used.  Also shown in 
the figure are the expected trigger rates from simulated data, for 
photo-electron thresholds of 50 and 60 per telescope. It can be seen that 
for all the cases, from single sector trigger rate to the entire array, 
experimental data is consistent with the trigger rate of about 55 
photo-electrons per telescope.

All the simulated events are again examined taking into account photo-electron 
threshold of 55 photo-electrons per telescope. Figure 4 shows the differential 
energy spectra for both $\gamma-$ray and proton events. The peaks of these 
distributions give the energy thresholds of the array. These are 900 GeV for 
$\gamma-$rays and 2250 GeV for protons. 

\begin{figure}[t]
\includegraphics[width=8.3cm]{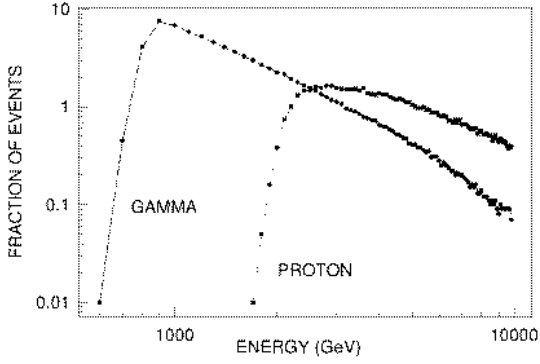}
\caption{Differential energy spectrum of triggered simulated events for 
$\gamma-$ray and proton initiated showers. The peak of the energy spectrum
or energy threshold is about 900 GeV for $\gamma-$rays.}
\end{figure}

\begin{figure}[t]
\includegraphics[width=8.3cm]{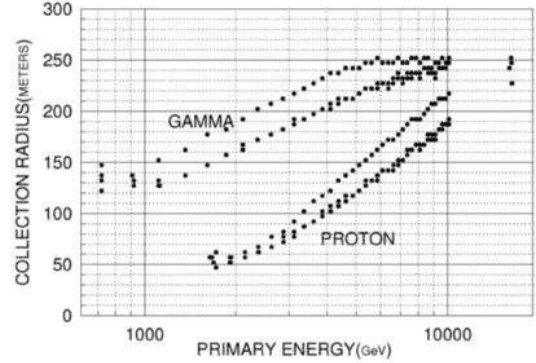}
\caption{Collection radius {\it vs} primary energy for showers initiated
by $\gamma-$rays and protons. For both the species lower curve is for 
single sector and upper one for entire array.}
\end{figure}

For each simulated event, the collection area is calculated. Figure 5 shows 
the collection radius for single and all the sectors of the array for 
showers initiated by $\gamma-$rays and protons. It should be noted that 
the relatively large collection radius (defined as the radius containing 
67\% of the events) of the PACT is essentially due to the large extent of 
the array. The saturation at large values of collection radius occurs 
because the events were generated only upto 300 m radius.

\section{Sensitivity of PACT}

Sensitivity of the experiment is minimum detectable flux of $\gamma-$rays
in presence of background of cosmic rays. It is related to the signal
to noise ratio given by 
\begin{equation}
{{S} \over {N}} \propto \sqrt{{{A_pT} \over {\Omega}}} E^{0.85-G}
\end{equation}
where $G$ is the $\gamma-$ray spectral index, $A_p$ the effective collection 
area, $T$ observation duration and $\Omega$ the solid angle of the telescope.


For PACT, $5\sigma$ sensitivity for an observation duration of 50 hours, 
above energy threshold of 1 TeV, is estimated to be $\sim$ 4.1 $\times$ 
10$^{-11}$ $ph~cm^{-2}~s{-1}$ for no background rejection. The angular 
resolution of PACT is estimated to be $\sim$ 0.1$^\circ$ (Majumdar 
{\it et al.}, 2001). This allows the rejection of about 99\% of off-axis 
background events using the arrival angle information alone. According to 
simulation studies, it is possible to reject a significant fraction of 
cosmic ray proton showers based on the information about arrival time of 
shower front at various telescopes (Chitnis and Bhat, 2001) and fluctuations
in density of \v Cerenkov photons at different telescopes (Bhat and Chitnis,
2001). We assume that at least 75\% of the background showers can be rejected 
using simulation based cuts. It may be mentioned here that the rejection 
efficiency of cosmic rays is much better than what is assumed here, as per 
simulation studies. However, we have not yet applied these cuts to the actual 
data. Hence we assumed an extremely conservative figure of hadron rejection 
efficiency. As a result, at least 99.75\% of the cosmic ray showers enetering 
the field of view could be rejected. Accordingly, the sensitivity improves 
to $\sim$ 4.1 $\times$ 10$^{-12}$ $ph~cm^{-2}~s^{-1}$. The efficiency of 
retaining $\gamma-$ray showers while exercising cuts for rejecting hadronic 
showers is estimated to be $\sim$ 44\%. Based on these conservative estimates, 
the minimum duration of observation required to detect Crab nebula at a 
significance level of 5$\sigma$ is $\approx$ 8 hours (4 hours ON source and 
4 hours OFF source). The sensitivity of PACT is compared with the present and 
future atmospheric \v Cerenkov experiments in Figure 6.

\begin{acknowledgements}
It is pleasure to thank A. I. D'Souza, P. J. Francis, S. R. Joshi, M. S. Pose, 
P. N. Purohit, K. K. Rao, S. K. Rao, S. K. Sharma, A. J. Stanislaus and 
P. V. Sudershanan for their participation in various aspects of the 
experiment.
\end{acknowledgements}

\end{document}